\documentclass[11pt,letterpaper]{article}
\usepackage{amsmath,bbm,amssymb,amsthm,graphics,epsfig,setspace,amsthm,hyperref,booktabs}
\usepackage[flushleft]{threeparttable}
\usepackage{natbib}
\usepackage{xcolor}

\usepackage{authblk}
\usepackage[normalem]{ulem}
 
\newtheorem{theorem}{Theorem}

\def\T{{ T }}
\def\xx{{\bf x}}
\def\uu{{\bf u}}
\def\bbeta{{\boldsymbol\beta}}
 
\def\be{\mathcal{B}}
\newcommand{\fv}{\mathcal{F}}
\newcommand{\re}{\mathbb{R}}

\DeclareMathOperator*{\argmin}{arg\,min}

\newcommand{\benu}{\begin{enumerate}}
	\newcommand{\eenu}{\end{enumerate}}
\newcommand{\bi}{\begin{itemize}}
	\newcommand{\ei}{\end{itemize}}

\begin{document}

\title{Interval-censored linear quantile regression} 

\author[1,2]{Taehwa Choi}
\author[3]{Seohyeon Park}
\author[4,5]{Hunyong Cho\thanks{This work was done outside of Amazon.com}}
\author[3]{Sangbum Choi\thanks{Corresponding author, 145 Anam-ro, Seongbuk-gu, Seoul 02841, South Korea. E-mail: choisang@korea.ac.kr}}

%\author{MMH}
\date{} 

\affil[1]{\small School of Mathematics, Statistics and Data Science, Sungshin Women’s University, Seoul 02844, South Korea} 
\affil[2]{\small Data Science Center, Sungshin Women’s University, Seoul 02844, South Korea}

\affil[3]{\small Department of Statistics, Korea University, Seoul 02841, South Korea} 

\affil[4]{\small Amazon.com, 
Seattle, WA 98109, U.S.A.} 
\affil[5]{\small Department of Statistical Science, 
Duke University, 
Durham, NC 27710, U.S.A.}

\maketitle

\begin{abstract}

Censored quantile regression has emerged as a prominent alternative to classical Cox's proportional hazards model or accelerated failure time model in both theoretical and applied statistics. 
While quantile regression has been extensively studied for right-censored survival data, methodologies for analyzing interval-censored data remain limited in the survival analysis literature. This paper introduces a novel local weighting approach for estimating linear censored quantile regression, specifically tailored to handle diverse forms of interval-censored survival data. The estimation equation and the corresponding convex objective function for the regression parameter can be constructed as a weighted average of quantile loss contributions at two interval endpoints. The weighting components are nonparametrically estimated using local kernel smoothing or ensemble machine learning techniques. To estimate the nonparametric distribution mass for interval-censored data, a modified EM algorithm for nonparametric maximum likelihood estimation is employed by introducing subject-specific latent Poisson variables. The proposed method's empirical performance is demonstrated through extensive simulation studies and real data analyses of two HIV/AIDS datasets.

\end{abstract}

\section{Introduction}
\label{sec1}

Censored quantile regression has emerged as a prominent alternative to classical Cox's proportional hazards model or accelerated failure time model in both theoretical and applied statistics \citep{po03,pe08,wa09}.
This approach allows for a comprehensive examination of the entire distribution of survival responses in relation to a group of covariates.
Censored quantile regression offers several advantages over traditional approaches to handling censored regression problems. Firstly, unlike methods that focus solely on modeling conditional means, it can differentiate effects across various conditional quantiles. Secondly, it enables consistent estimation in censored regression models with significantly fewer distributional assumptions than typically required. Thirdly, it retains the benefits associated with quantile regression, including robustness and a comprehensive analytical approach. Furthermore, in fields like economics, political science, psychology, sociology, and biostatistics, it has been increasingly utilized to unearth potential predictive relationships between variables, particularly in scenarios where the relationships between their means are non-existent or weak.

The estimation problem of quantile regression can be reformulated as a linear programming problem, to which  simplex or interior point methods can be applied  \citep{koenker05}. 
When data are subject to censoring, however, statistical estimation and inference for quantile parameters are much more involved.
\cite{po84} first studied censored quantile regression with fixed censoring.  For independent random right-censoring, \cite{ying1995survival,bang02} proposed a semiparametric median regression model,
which was later extended to competing risks quantile regression \citep{peng2009competing,ch18}.
\cite{po03} and \cite{wa09} developed a novel weighted estimating approach for right-censored data, motivated by classical and local Kaplan–Meier estimators, respectively. Using the martingale representation, \cite{pe08} studied another regression quantile  estimator motivated by the Nelson–Aalen estimator. More recently, \cite{ya18} developed a general multiple-imputation method, based on a variation of the data augmentation algorithm, and \cite{de2019adapted} developed a majorize-minimize (MM) algorithm to directly minimize an adaptive censored quantile loss function. \cite{lee2022efficient} studied the connection between \cite{pe08}'s and \cite{wa09}'s estimators and provided a unified inferential method that covers  conventional approaches as special cases.
\cite{son2022quantile} proposed quantile regression for competing risks data from stratified case-cohort studies.

Although quantile regression has been widely studied for right-censored survival data, methodologies for analyzing interval-censored data remain scarce in the survival analysis literature.
The term `interval-censoring' refers to situations where the exact time of an event of interest is not precisely known, but rather it is identified as occurring within a specific time interval.
This type of sampling scheme occurs, for example, when patients are periodically  assessed for disease recurrence or progression.
\cite{ou16} proposed current status quantile model by approximating the objective function to the difference of two hinge loss functions, which is, however, is computationally demanding due to a complex grid-search algorithm for parameter estimation. 
\cite{kim10ic} constructed an estimating equation based on weights derived from interval-censored data.
The quantile regression estimator by \cite{zh17} is based on quantile order-deterministic cases only and thus statistically inefficient. \cite{fru22} introduced a procedure based on local-weighted estimating equations for interval-censored data. 
\cite{li12} and \cite{ji12} generalized right-censored quantile methods  to doubly-censored data, based on the idea of redistribution-of-mass \citep{po03} and martingale equation \citep{pe08}, respectively.

In this article, we propose a novel 
local-weighting approach for estimating linear censored quantile regression and present its application to  various types of interval-censored survival data.
The estimating equation and the corresponding convex objective function for the regression parameter can be  constructed as a weighted average of quantile loss contributions at two interval endpoints.
This formulation leverages the efficient quantile check function  for uncensored data, while for the interval-censored observations, the nonparametric weights  adjust for the redistribution of mass due to interval-censoring. 
It should be noted that we employ an estimating function analogous to the one described in \cite{fru22}. However, there is a notable methodological difference: \cite{fru22} used a parametric piecewise-constant hazard regression for modeling the residual distribution given covariates. In contrast, our approach centers on estimating the nonparametric residual distribution function for covariates under interval-censoring. To achieve this, we propose two methods for computing the nonparametric distribution mass in interval-censored data: local kernel smoothing and ensemble machine learning techniques.

One approach to implementing local kernel smoothing is to solve the self-consistency equation for interval-censored data \citep{tu76}. However, this method does not account for the effect of covariates on the error distribution and may not characterize nonparametric maximum likelihood estimation (NPMLE) due to the non-existence or non-uniqueness of the self-consistency solution \citep{we97}. Instead, we use a modified expectation-maximization (EM) algorithm for nonparametric likelihood estimation, based on subject-specific latent Poisson variables \citep{ze16,ch21}, which allows us to reconstruct the local likelihood function in a standard survival analysis context.
To enhance the flexibility in modeling the weighting components, our approach incorporates a nonparametric recursive partitioning ensemble method known as interval-censored recursive forests (ICRF), as detailed in \cite{cho2022interval}. This nonparametric regression estimator addresses the splitting bias problem of existing tree-based methods and iteratively updates survival estimates in a self-consistent manner. A notable advantage of the ICRF method over local kernel smoothing techniques is its enhanced prediction accuracy, particularly in scenarios with a moderate-to-high dimensionality of covariates. The effectiveness of the proposed methods for interval-censored quantile regression is empirically validated through extensive simulation studies and the analysis of two HIV/AIDS datasets.

The remainder of this article is organized as follows. 
In Section \ref{sec2}, we introduce our locally weighted quantile regression method for interval-censored data.
{In Section \ref{sec:np}, we describe} the estimating methods for conditional distribution function based on kernel-smoothing-weighted EM algorithm and recursive partitioning ensemble approach.
Section \ref{sec3} presents extensive simulations that demonstrate the finite sample performance of the methods, while
Section \ref{sec4} provides two HIV-data examples to illustrate the proposed methods. We conclude with the final discussion and  remarks in Section \ref{sec5}.
A sample code for implementing our method can be found on the first author's GitHub repository (\url{https://github.com/taehwa015/ICQR}).

\section{Interval-censored quantile regression}
\label{sec2}

\subsection{Model and data}

Let $T$ be the log-transformed failure time, and $\xx$ be a $p$-dimensional vector of time-independent covariates, the first element of which is set to one for intercept.
Our main  objective is to estimate the $p$-dimensional quantile coefficient vector $\bbeta_0(\tau)$ for a fixed $\tau\in(0,1)$ in the linear quantile regression model 
\begin{equation}
	T=\xx^\T\bbeta_0(\tau) +e(\tau)
	\label{cqr}
\end{equation} 
under a random interval-censored sampling scheme,
where $e(\tau)$ is an independent error with its $\tau$th quantile being zero.
If the quantile assumption on $e(\tau)$ is replaced by $E[e(\tau)]=0$, this model corresponds to the familiar accelerated failure time (AFT) model \citep{ji03,choi2021fast}.  
Let denote the covariate-conditional $\tau$th quantile function of $T$ by $Q_{T}(\tau|\xx) = \inf \{t : F(t|\xx) \ge \tau \}$, where $F(\cdot|\xx)$ represents the cumulative distribution function of $T$ conditional on $\xx$.
Then, based on an i.i.d. sample $(T_i,\xx_i),i=1,\ldots,n$, from $(T,\xx)$, model \eqref{cqr} implies
$Q_{T}(\tau | \xx_i) = \xx_i^T \bbeta_0(\tau )$,
which suggests a new estimation strategy 
that differs from conventional mean-based approaches to analyze survival data.

Unlike classical regression models for survival data, such as Cox's proportional hazards model and AFT model, model \eqref{cqr} allows for quantifying heteroskedasticity of the data by evaluating the covariate effect at different quantile levels. 
In the uncensored case, it is well known that the estimation of $\bbeta_0(\tau)$ can be done by minimizing the objective function
\begin{equation}\label{check}
\tilde S_n(\bbeta) = n^{-1} \sum_{i=1}^{n} \rho_\tau(T_i - \xx_i^\T\bbeta),
\end{equation} 
where $ \rho_\tau(u)  =  u\{ \tau - I(u \le0) \}$ is the quantile loss function, or equivalently by solving the  estimating equation
\begin{equation} \label{qee}
\tilde M_n(\bbeta) = n^{-1} \sum_{i=1}^{n} \xx_i \{ \tau - I(T_i - \xx_i^\T\bbeta \le 0)  \} \approx 0.
\end{equation}
For random right censoring, the main rationale of the current literature has been so far to take censoring into account through the formulation of synthetic data points or weighting schemes.  
See \cite{po03,pe08,wa09}, among others, for quantile regression analysis of right-censored survival data.

Let us now formulate the mixed-case interval-censoring by assuming that there may exist a random sequence of {log-transformed} examination times for each individual, denoted by
$-\infty= U_0 <U_1< U_2 <\cdots<U_K < U_{K+1}=\infty$,
where $K$ represents the number of random examination times.
Let $\Delta$ denote the interval-censoring indicator that takes 1 or 0 if $T$ is exactly observed or interval-censored. For the case of interval-censoring (i.e., $\Delta=0$), we may identify the tightest interval $(L,R)$ that contains the unknown failure time $T\in(L,R)$, such that 
$L= \max \{ U_k : U_k \le T,~k=0,\ldots,K \}$ and
$R=\min \{ U_k : U_k>T,~k=1,\ldots,K+1 \}$, 
and if $\Delta=1$, we let $L=R=T$. 
Then, the ``effective'' observed data from a random sample of $n$ subjects consist of i.i.d. copies of $(\Delta T,(1-\Delta)L,(1-\Delta )R,\Delta,\xx)$, that is, 
\begin{equation} \label{ic-data}
\{ (\Delta_iT_i, (1-\Delta_i)L_i, (1-\Delta_i)R_i,\Delta_i, \xx_i) : ~i=1,\ldots,n\}.
\end{equation}

In this setup, it is assumed that the joint distribution of $(U_1,\ldots , U_K)$ is independent of $T$ given $\xx$.
When the data are interval-censored, it suffices to know the values of $(L,R)$, since the other examination times do not contribute to the quantile estimation \citep{ze16}.
In the statistical literature, this type of data is often called partly interval-censored (PIC) data \citep{gao17} if $P(\Delta=1)>0$. If $P(\Delta=1)=0$, it is reduced to ``case-1'' and ``case-2'' interval-censored data, when $K=1$ and $K\ge2$, respectively.
Note that $L = -\infty$ and $R=\infty$ represents left- and right-censored observation, respectively, and 
if it is either $L=-\infty$ or $R=\infty$, (\ref{ic-data}) reduces to doubly-censored (DC) data \citep{ch21}. 
Therefore, our data configuration is general enough to cover various forms of interval-censored data. In the following, we shall focus on the PIC sampling scheme, as it can include other interval-censoring types as  special cases.

\subsection{Estimating procedure}

Under the general interval-censoring sampling scheme described above, we can observe the exact failure time $T_i$ only when $\Delta_i=1$, and otherwise the time interval $(L_i,R_i)$ that contains $T_i$. 
In this context, we propose to estimate $\bbeta_0(\tau)$ by minimizing the following  convex interval-censored quantile loss function 
\begin{equation}\label{obj}
\begin{split}
	S_n(\bbeta,F )=n^{-1}\sum_{i=1}^n
	& \, \Big[\Delta_i\rho_\tau(T_i-\xx_i^\T \bbeta)+ (1-\Delta_i)\times \\
	& \quad \{w_i(F )\rho_\tau(L_i-\xx_i^\T \bbeta)+
	(1-w_i(F ))\rho_\tau(R_i-\xx_i^\T \bbeta)\}\Big],
\end{split}	
\end{equation} 
where  the weight function $w_i(F)$ for the individuals with interval-censored observations
(i.e., $\Delta_i=0$) is specified as 
\begin{align}\label{wt}
		w_i(F )= 
		\begin{cases} 
		1, & F (L_i|\xx_i) \ge \tau,~\Delta_i=0 ;\\ 
		\dfrac{\tau-F (L_i|\xx_i)}{F (R_i|\xx_i)-F (L_i|\xx_i)},& F (L_i|\xx_i) < \tau< F (R_i|\xx_i),~\Delta_i=0 ;\\
		0, &F (R_i|\xx_i)\le\tau,~\Delta_i=0. 
			\end{cases} 
\end{align}

Since the negative subgradient of $S_n(\bbeta,F )$ about $\bbeta(\tau)$ is equal to 
\begin{equation}\label{ee}
\begin{split}
			M_n(\bbeta,F )=n^{-1}\sum_{i=1}^n
		&  \xx_i\Big[\tau-\Delta_iI(T_i-\xx_i^\T \bbeta\le0)- (1-\Delta_i)\times \\
		& \quad \{w_i(F )I(L_i-\xx_i^\T \bbeta\le0)+
		(1-w_i(F ))I(R_i-\xx_i^\T \bbeta\le0)\}\Big],
\end{split}	
\end{equation}
the quantile estimator $\hat\bbeta(\tau)$ would be obtained 
by minimizing the objective function \eqref{obj} or equivalently solving the estimating equation \eqref{ee} about $\bbeta$
{given a fixed underlying error distribution $F$}.
The proposed weighting method in \eqref{wt} extends the idea of \cite{wa09} to general interval-censored sampling cases, based on the argument of the redistribution-of-mass for censored observations \citep{ef67}. 
As the estimating equation in \eqref{qee} suggests, 
the contribution of each observation for estimating $\bbeta_0(\tau)$ 
only depends on the sign of the residual $e_i(\bbeta) = T_i - \xx_i^\T\bbeta$ in the uncensored case given $\bbeta$. 
Notice that, when $\Delta_i=1$, our estimating function \eqref{ee} is equivalent to  equation \eqref{qee}.
When $\Delta_i=0$, however, the sign of $e_i(\bbeta)$ is not fully determinate based on the observed data, because we only know that $L_i< T_i<R_i$ but cannot exactly locate $T_i$. Our approach to this case is to take a proper weighted average between $I(L_i- \xx_i^\T\bbeta\le0)$ and $I(R_i- \xx_i^\T\bbeta\le0)$ as shown in equation \eqref{ee}.

To gain further insight into our proposed estimating procedure and understand how interval-censoring is accounted for through the weighting term  \eqref{wt}, we examine three separate cases one by one: 
(i) $F (L_i|\xx_i) \ge \tau,~\Delta_i=0$;
(ii) $F (L_i|\xx_i) < \tau< F (R_i|\xx_i),~\Delta_i=0$ ;
and 
(iii) $F (R_i|\xx_i)\le\tau,~\Delta_i=0$. 
Case (i) means that {$\xx_i^T\bbeta\le L_i<T_i$}, implying that $I(T_i-\xx_i^T\bbeta\le 0)=0$, for which we assign $w_i(F)=1$ because $I(T_i-\xx_i^T\bbeta\le0)=I(L_i-\xx_i^T\bbeta\le0)=0$. 
Similarly, since {$T_i<R_i\le\xx_i^T\bbeta$} under case (iii), we immediately know that $I(T_i-\xx_i^T\bbeta\le0)=I(R_i-\xx_i^T\bbeta\le 0)=1$ and thus let 
$w_i(F)=0$. The ambiguous situation is case (ii), for which 
$L_i<\xx_i^T\bbeta<R_i$, and $L_i<T_i<R_i$.
Although it is clear that 
$I(L_i-\xx_i^T\bbeta\le0)=1$ and 
$I(R_i-\xx_i^T\bbeta\le0)=0$,
the order between $T_i$ and $\xx_i^T\bbeta$ is indeterminate.
In this case, we leverage the fact that
\begin{align}\label{ewt}
\begin{split}
    	E\{ I(T_i - \xx_i^\T\bbeta_0(\tau)\le0) | L_i< T_i<R_i ,\xx_i\}  
 &= \dfrac{P(L_i < T_i \le \xx_i^\T\bbeta_0(\tau)|\xx_i)}{P(L_i < T_i <R_i|\xx_i)} \\
	&= \dfrac{\tau - F (L_i|\xx_i)}{ F (R_i|\xx_i) - F (L_i|\xx_i)}.
\end{split}
\end{align}
Therefore, by assigning the weight $w_i(F)=\frac{\tau-F(L_i|\xx_i)}{F(R_i|\xx_i)-F(L_i|\xx_i)}$
to the ``pseudo observation'' at the left endpoint $L_i$ and 
redistributing the complementary weight $1-w_i(F)$ to the right endpoint $R_i$, the estimating function \eqref{ee} yields unbiased results. 
This is true because when $\Delta_i=0$ the following equality holds in any of the three scenarios:
$$
E\{w_i(F )I(L_i-\xx_i^\T \bbeta_0\le0)+
(1-w_i(F ))I(R_i-\xx_i^\T \bbeta_0\le0)|\Delta_i=0,\xx_i\}
=	E\{ I(T_i - \xx_i^\T\bbeta_0\le0) | \Delta_i=0 ,\xx_i\}.
$$

Intuitively, our approach mixes the quantile contributions of the two data points $L_i$ and $R_i$ in proportion to their relative distance from $\tau$ in percentile.
From the above-described reasoning, it also naturally appears that our general weight function  \eqref{wt} may then be extended to left censoring and right censoring. 
If subject $i$ is left-censored
(i.e., $L_i=-\infty$), 
$
w_i(F)=	E\{ I(T_i - \xx_i^\T\bbeta_0(\tau)\le0) |T_i<R_i ,\xx_i\} 
=\frac{\tau }{ F (R_i|\xx_i) },
$
and if it is right-censored (i.e., $R_i=\infty$),
$
w_i(F)=	E\{ I(T_i - \xx_i^\T\bbeta_0(\tau)\le0) |T_i>L_i ,\xx_i\} 
=\frac{\tau-F(L_i|\xx_i) }{1- F (L_i|\xx_i) }.
$
Hence, our weighting method can also accommodate DC data, which may include
left censoring and right censoring as special cases of interval-censoring.

\subsection{Implementation}
 
Clearly, our proposed estimating procedure requires a consistent nonparametric estimator of the covariate-conditional error distribution function $F(\cdot|\xx)$, for which we propose to consider  local self-consistent nonparametric maximum likelihood (NPML) and interval-censored recursive forests (ICRF) estimators. These methods will be discussed more thoroughly in Section 3. 
Accordingly, given a consistent estimator $\hat F$ of $F$,  we can define the proposed quantile regression estimator $\hat\bbeta(\tau)$ as the minimizer of $S_n(\bbeta,\hat F)$, i.e., 
$$
\hat\bbeta(\tau)=\argmin_{\bbeta\in\mathbb{B}}
S_n(\bbeta,\hat F),
$$
where 
\begin{align}\label{Snhat}
\begin{split}
	S_n(\bbeta,\hat F)=n^{-1}\sum_{i=1}^n
	& \, \Big[\Delta_i\rho_\tau(T_i-\xx_i^\T \bbeta)+ (1-\Delta_i)\times \\
	& \quad \{w_i(\hat F)\rho_\tau(L_i-\xx_i^\T \bbeta)+
	(1-w_i(\hat F))\rho_\tau(R_i-\xx_i^\T \bbeta)\}\Big].
\end{split}	
\end{align}
Note that $S_n(\bbeta,\hat F)$ exploits all observations in the estimating process, regardless of their censoring status.
On the other hand, the estimation procedure of \cite{zh17} uses only the information from the order-deterministic cases, i.e.,  cases (i) and (iii), while assigning zero weights to the order-indeterministic observations from case (ii).  
Their methodology seems to lead to suboptimal performance, characterized by significant biases, particularly when dealing with small-to-moderate sample sizes and scenarios where the examination times are coarse.

The proposed locally weighted interval-censored quantile regression is simple to implement with currently available \texttt{R} software.
To this end, we may create an augmented ``pseudo'' dataset by appending 
$ \{ (L_i, \xx_i)\}_{\Delta_i=0} $
and 
$ \{ (R_i, \xx_i) \}_{\Delta_i=0}$
to $\{(T_i,\xx_i)\}_{\Delta_i=1}$.
For left-censored and right-censored observations, we may set  
$ L_i = -M^*  $ and $ R_i = M^*$, respectively, where a constant $M^*>0$ should be large enough to ensure $M^*>|\xx_i^T\bbeta|$ for any combination of $\xx_i$ and $\bbeta$. 
To compute $\hat\bbeta$, we then apply the \texttt{rq()} function from the \texttt{R:quantreg} package \citep{ko08} to the augmented dataset with corresponding local weights, i.e., 
1 to $\{(T_i,\xx_i)\}_{\Delta_i=1}$, 
$w_i(\hat F)$ to 
$ \{ (L_i, \xx_i)\}_{\Delta_i=0} $,
and 
$1-w_i(\hat F)$ to 
$ \{ (R_i, \xx_i)\}_{\Delta_i=0} $,
respectively.

On the other hand, statistical inference about $\bbeta_0(\tau)$  is not straightforward, 
because $M_n(\bbeta,F)$ is a non-smooth function about $(\bbeta,F)$ and the asymptotic behavior of the proposed nonparametric estimator $\hat F$ is too complicated and not fully known except for its consistency. 
To make inferences about {$\bbeta_0(\tau)$}, we use 
the perturbed resampling method \citep{ji01}. By introducing the perturbation random variable, we can derive perturbed nonparametric estimators and quantile estimators through the construction of perturbed nonparametric likelihood function and objective function, respectively.
The 95\% percentile bootstrap confidence interval for $\bbeta_0(\tau)$ 
can be obtained from the 2.5th and 97.5th percentiles of bootstrapped coefficients or by using the normal approximations of bootstrapped samples.
The detailed procedure is described in the web-based appendix.

\subsection{Asymptotic results}

In this section, we show that the proposed quantile regression estimator $\hat\bbeta(\tau)$
is consistent for $\bbeta_0(\tau)$.
We first impose the following regularity conditions.
\begin{description}
	\item[\rm (C1)] 
	The parameter space $\be$ containing $\bbeta_0$ is a known compact set in $\re^p$. 
	For the covariate vector $\xx\in\mathcal{X} $,
	there exists a constant $K_0>0$ such that $E(\|\xx\|^2)\le K_0$, $\max_{1\le i \le n} \|\xx_i\| = O_p(n^{1/2}(\log n)^{-1})$,
	and $E(\xx\xx^\T)$ is  positive definite.
	\item[\rm (C2)]
    If $F_0(\cdot|\xx)$ is the distribution function of $T$ conditional on $\xx\in\mathcal{X}$, 
    it is uniformly bounded away from zero with respect to $t\in [a,b]$ and $\xx$ 
    and has  a strictly positive and continuously differentiable density function $f_0(\cdot|\xx)$ on $[a,b]$, where $[a,b]$ is the union of the support of $(L,R)$.
	\item[\rm (C3)] 
	The joint distribution of $(L,R)$ given $\xx$, denoted by $G(l,r|\xx)=P(L\le l, R\le r|\xx)$, is uniformly bounded away from zero with respect to $(l,r) \in [a,b]$ and $\xx$,
    and has a strictly positive and continuously differentiable density function $g(\cdot,\cdot|\xx)$ on the support.
    Their marginal distribution functions $G_L(\cdot|\xx) = \int g(\cdot,r|\xx)dr$
	and $G_R(\cdot|\xx)=\int g(l,\cdot|\xx)dl$ are absolutely continuous on $[a,b]$ if $0 < G_L(\cdot|\xx) < G_R(\cdot|\xx)<1$.
    \item[\rm (C4)] 
 The kernel function $K(\cdot)>0$ has a compact support.
 It is Lipschitz continuous of order 1, and satisfies 
 $\int K(u)du = 1, \int u K(u)du =0, \int K^2(u)du <\infty$, and 
 $\int |u|^2K(u)du <\infty$.
 The bandwidth $h_n \in \re^+$ satisfies $nh_n / \log n \to \infty$ and $\log(1/h_n) / \log\log n \to\infty$ when $h_n \to 0$ and $h_n/h_{2n}$ is bounded.
\end{description}
Condition (C1) is a classical assumption in linear regression that the parameter space $\be$ is a compact space in $\re^p$ 
and the covariate vector $\xx$ is uniformly bounded in $\mathcal{X}$.
(C2) and (C3) state the uniform boundedness and sufficiently smoothness conditions 
for the distribution function $F\in\fv$ and density functions of the examination interval $(L,R)$, respectively.
Condition (C4) is required for the kernel function and the bandwidth, which is regularly assumed in kernel-based nonparametric estimation \citep{einmahl05}.
\begin{theorem}\label{thm1}
	Under the conditions (C1)--(C4),
	the quantile regression estimator $\hat\bbeta(\tau)$ satisfies that $ \hat\bbeta(\tau)  \to \bbeta_0(\tau) $
	in probability as $n\to\infty$.
\end{theorem}
Theorem \ref{thm1} can be proved by checking the conditions in the Theorem 1 of \cite{chen03}, which discusses the asymptotic results of M-estimators when the criterion function is not sufficiently smooth.
We provide the detailed proof in the web-based supplementary note.

\section{Nonparametric redistribution-of-mass of interval-censoring}
\label{sec:np}

\subsection{Kernel-based maximum likelihood estimation}

Now we propose a nonparametric multivariate kernel-based approach to estimate the conditional distribution function $F(\cdot|\xx)$ or equivalently the corresponding cumulative hazards function $\Lambda(\cdot|\xx) = -\log \{  1- F(\cdot|\xx)\} $. 
It is assumed that $\Lambda(t|\xx)=\int_{-\infty}^t\lambda(s|\xx)ds$ 
is continuous and at least thrice differentiable,
where $\lambda(t|\xx)$ is a conditional hazard function at time $t$ given $\xx$. 
To implement local maximum likelihood estimation, however, we shall regard $\Lambda(\cdot|\xx)$ as a non-decreasing step function $\Lambda(\cdot;\xx)$ with distinct jumps at unique values of observed event times. 
Let  $-\infty=s_0 < s_1 < \cdots < s_m < \infty$ denote the unique and ordered event times of the observed set of {$\{\Delta_{i}T_i, (1-\Delta_{i})I(L_i>-\infty)L_i, (1-\Delta_{i})I(R_i<\infty)R_i:i=1,\ldots,n\}$},
where $m$ is the number of distinct time points in a finite time horizon.
With a slight abuse of notation ({with $\xx$ being fixed}), we may write $d\Lambda_j\equiv d \Lambda(s_j;\xx)$ to denote the jump size of $\Lambda(\cdot;\xx)$ at time $s_j$ and let $\Lambda_j=\Lambda(s_j;\xx)\equiv \sum_{t\le s_j} d\Lambda(t;\xx)$. We assume that $\Lambda(-\infty|\xx)=0$ and $\Lambda(\infty|\xx)=\infty$.

We can first formulate the nonparametric local  log-likelihood function for $\Lambda(\cdot|\xx)$  as 
\begin{equation}
\begin{split}
	\ell(\Lambda|\xx) = n^{-1} \sum_{i=1}^n B_{ni}(\xx) \Bigg(
	&\Delta_i \Bigg\{ \sum_{j=1}^m I(s_j = T_i)\log \lambda(T_i|\xx) -\Lambda(T_i|\xx) \Bigg\}+  \\
	&(1-\Delta_i)\times\log\Big[ \exp\{ -\Lambda(L_i|\xx) \} - \exp\{-\Lambda(R_i|\xx)\}\Big]\Bigg),
\end{split}
\end{equation}
where $\{B_{ni}(\xx)\}_{i=1}^n$ denotes a sequence of non-negative weights, scaled to sum up to one.
By replacing $\lambda(\cdot|\xx)$ and $\Lambda(\cdot|\xx)$ with their discrete versions, $d\Lambda(\cdot;\xx)$ and $\Lambda(\cdot;\xx)$, respectively, the above log-likelihood formulation can be alternatively written as 
\begin{align}\label{like}
	\begin{split}
	\ell(\Lambda;\xx)=	n^{-1} \sum_{i=1}^n B_{ni}(\xx) &\Bigg[
		\Delta_i \Bigg\{ \sum_{j=1}^m I(s_j = T_i)\log d\Lambda_j -\sum_{j:s_j\le T_i} d\Lambda_j\Bigg\} +(1-\Delta_i)\times \\
		&\log\bigg\{ \exp\Bigg( -\sum_{j:s_j\le L_i} d\Lambda_j\Bigg) - \exp\Bigg( -\sum_{j:s_j\le R_i} d\Lambda_j\Bigg)\Bigg\}\Bigg].
	\end{split}
\end{align}

For the local weight $B_{ni}(\xx)$, we may use the Nadaraya-Watson's method 
$$
B_{ni}(\xx)=\frac{K\Big(\frac{\xx-\xx_i}{h_n}\Big)}{\sum_{j=1}^n K\Big(\frac{\xx-\xx_j}{h_n}\Big)}, ~~i=1,\ldots,n,
$$ 
where $K(\cdot)$ is a kernel density function and $ h_n \in  (0,\infty) $ is a bandwidth converging to 0 as $ n \to\infty$.
With multiple covariates $\xx\in\mathbb{R}^p~(p>1)$, we may consider a multivariate product kernel 
{$ K(\uu) = \prod_{k=1}^{p} k(u_k)  $} for $\uu= (u_1,\ldots,u_p)^\T$, which works well with a small number of covariates. 
In our implementation, we used a Gaussian kernel for {$k(u),u\in\mathbb{R}$} for convenience because kernel selection generally does not make a significant difference in estimation performance.
On the other hand, it is well recognized that the choice of bandwidth is crucial to kernel-based methods because a non-optimal bandwidth may lead to suboptimal performance. Our experience reveals that the proposed local weighting method is quite robust to the choice of bandwidth in the setting of  quantile regression. Hence we use the conventional normal scale bandwidth selection  method \citep{sil86}, for which 
$h_n = O_p(n^{-1/5})$ follows by assuming that the error is normally distributed.

The local log-likelihood function \eqref{like} can be optimized in different ways, for example, by directly solving the score equation of $\ell(\Lambda; \xx)$ {for} each $d\Lambda_j$ \citep{ch20} or 
employing an EM algorithm that takes the advantage of the fact that the  likelihood function of interval-censored data can be understood as the likelihood of a latent Poisson distribution \citep{ze16,ch21}. 
We appeal to the latter approach, because it is simple, intuitive, and ensures that the solution is an NPMLE. 
To invoke the EM algorithm, let $\{W_{ij}\}_{j=1}^m$ be independent subject-specific Poisson random variables with mean $\mu_j=d\Lambda_j$ with  density function $ p(W = w; \mu) = e^{-\mu} \mu^{w} /w!$. 

Then the log-likelihood function \eqref{like} can be equivalently expressed as 
\begin{equation*} 
\begin{split}
	n^{-1} & \sum_{i=1}^n B_{ni}(\xx)  \Bigg( \Delta_{i} \Bigg\{\sum_{j=1}^m I(s_j=T_i) \log P(W_{ij}=1) + \log P\Bigg(\sum_{j:s_j\le T_i} W_{ij}=0 \Bigg)\Bigg\} \\
	&+(1-\Delta_{i}) \Bigg[ \log P\Bigg(\sum_{j:s_j \le L_i}W_{ij} = 0\Bigg) +  \log \Bigg\{1- P\Bigg(\sum_{j:L_i< s_j \le R_i}W_{ij} > 0\Bigg)\Bigg\}\Bigg]\Bigg).
\end{split}
\end{equation*}

We maximize the above expression through an EM algorithm by treating $W_{ij}$ as missing data.
Assuming that the $W_{ij}$'s are known, the pseudo complete-data log-likelihood function takes a simple form
$$
\ell^c(\Lambda;W) = n^{-1}\sum_{i=1}^n\sum_{j=1}^m  B_{ni}(\xx) p(W_{ij},d\Lambda_j) ,
$$
under the constraints that 
(i) if $\Delta_i = 1$, $W_{ij} = 1$ at $s_j = T_i$ and
 $\sum_{j:s_j<  T_i} W_{ij} = 0$,
(ii) if $\Delta_i = 0$, $\sum_{j:s_j\le L_i} W_{ij} = 0$ and $\sum_{j:L_i<s_j\le R_i }W_{ij} > 0$. 
After some algebra for the E-step of the EM algorithm, we can show that the expected  complete-data log-likelihood function is equivalent to 
\begin{align}\label{plike}
	E_W \{\ell^c(\Lambda; W,\xx)\} =\sum_{i=1}^n 
	B_{ni}(\xx)  \sum_{j=1}^m   I(s_j \le \tilde T_i)
	\{ \log (d\Lambda_j) \xi_{ij}-d\Lambda_j \}
\end{align}
by ignoring some constant terms, 
where $\tilde T_i = \Delta_iT_i + (1-\Delta_i) \{R_i I(R_i<\infty) + L_i I(R_i=\infty)\}$ and
the conditional expectation $\xi_{ij}\equiv E_W(W_{ij})$ of $W_{ij}$ under the aforementioned constraints is given by 
\begin{align} \label{e-step}
	\begin{split}
		 \xi_{ij}= \Delta_i I(s_j = T_i) 
		+(1-\Delta_i)\Bigg\{ \dfrac{d \Lambda_j I(L_i < s_j \le R_i) }{1 - \exp(-\sum_{j:L_i<s_j\le R_i}d\Lambda_j )} 
		\Bigg\} + d\Lambda_j I(s_j > \tilde T_i).
	\end{split}
\end{align}
Notice that  the complete-data log-likelihood function in \eqref{plike} resembles standard log-likelihood function for right-censored survival data. Therefore, in the M-step, we can obtain a closed-form expression for $d\Lambda_j$ by solving the score likelihood equation of \eqref{plike} with respect to $d\Lambda_j$, which leads to 
\begin{align}\label{m-step}
	d  \hat\Lambda_{j} =\frac{\sum_{i=1}^nB_{ni}(\xx)\xi_{ij} I(s_j \le \tilde T_i) }
	{\sum_{i=1}^n B_{ni}(\xx)I(s_j \le \tilde T_i)},~~j=1,\ldots,m.
\end{align}

To apply the EM algorithm, we first initialize each value of  $d\hat\Lambda_j$, such as $d\hat\Lambda_j^{(0)}=1/m$. 
The proposed EM-based nonparametric maximum likelihood estimator for $\Lambda(\cdot|\xx)$ can then be  calculated by simply iterating E-step \eqref{e-step} and M-step \eqref{m-step} until convergence. Our procedure for updating  $\hat\Lambda^{(k)}$ at the $k$th step is set to stop when either $\|\hat\Lambda^{(k)}-\hat\Lambda^{(k-1)}\|\le 10^{-5}$ or the maximum number of iterations (set to 100) is first achieved. We found that the computational cost for this operation is very small as it usually converges within 5-10  iterations. 
The resulting local maximum likelihood estimator for $F(\cdot|\xx)$ is then obtained by $ \hat F(\cdot|\xx) = 1-\exp\{-\hat\Lambda(\cdot|\xx) \} $.

\subsection{Interval-censored random forests}

Nonparametric multivariate kernel-based estimators usually suffer a great deal from the well-known curse of dimensionality.
Moreover, the kernel-smoothing method requires all covariates to be continuous and encounters practical difficulties even with  a moderate number of covariates. In this situation, one may use alternatively a parametric regression approach \citep{fru22} or a semiparametric proportional hazards model to characterize the covariate-conditional distribution distribution. However, these methods require a specific form of hazard structure between failure time and covariates, such as proportionality. 
As a practical alternative, we consider a new weighting approach that uses nonparametric recursive partitioning of interval-censored data \citep{cho2022interval}, i.e., interval-censored recursive forests (ICRF).

ICRF fits a random forest using ``extremely randomized trees" (ERT, \cite{geurts2006}) algorithm for each iteration. ERT is one of many tree-ensemble methods, where different trees are grown and then averaged. 
Unlike the prototypical random forests \citep{breiman2001}, which use subsampling for each tree and random variable selection for splitting at each intermediate node, ERT does not use the bootstrap subsampling but considers a randomly selected cutoff for each candidate variable at each node. 
ICRF also assumes conditional independent censoring between $T$ and the underlying visit process $\{U_k\}$ given $\xx$ unlike other tree-based methods; while the log-rank test is often used as a splitting rule in survival random forests, it assumes independent censoring across two subgroups. ICRF addresses this issue by using an extended version of the test, setting a self-consistency equation \citep{ef67}, and solving it through recursion.

The algorithm of ICRF solves for the covariate-conditional survival probabilities through the following self-consistent equation $$
S(t|\xx) = \frac 1 {n_{\text{tree}}} \sum_{b=1}^{n_{\text{tree}}} \frac 1 n\sum_{i=1}^n S(t|I_i, \xx_i \in A_b(\xx; S))\frac{1(\xx_i \in A_b(\xx; S))}{|A_b(\xx; S)|/n},
$$ 
where $I_i = (L_i,R_i]$, $A_b(\xx;S)$ is a terminal node of the $b$th tree that contains $\xx$, $|\cdot|$ is the sample size of the node, $n_{\text{tree}}$ is the number of trees in a forest. 
The estimated survival probability function obtained in the previous iteration is used to provide insight into the conditional survival probability of each subject by projecting the estimate function onto each subject's interval. This process can naturally handle covariate-driven informative censoring and provide richer information accumulated through recursion.

In ICRF, two splitting rules were developed to deal with interval-censoring: the extended Wilcoxon's rank-sum and log-rank tests. Both statistics could be interpreted as averages of the original test statistics of random realizations of the event time conditional on  censored intervals, where the conditional distribution is given by the survival probability estimates at the previous iteration.
A tree is grown by repeating the splitting procedure until each terminal node is either homogeneous enough or any terminal node does not become smaller than some fixed value. Finally, for each terminal node, the node-conditional survival probability is obtained by NPMLE. The tree estimators are averaged into a random forest estimator and kernel smoothing may be additionally applied to alleviate the discreteness along the time domain in interval-censoring.
Since the self-consistency algorithm or recursion does not guarantee convergence in general, ICRF modifies the ERT algorithm by holding out a small portion (5\%) of the sample for convergence monitoring. ICRF then returns the random forest of the iteration with the smallest error, where the error is calculated based on the hold-out sample.
The ICRF estimator satisfies the uniform consistency under some regularity conditions \citep{cho2022interval}. 
The implementation of ICRF is supported by the \texttt{icrf()} function in the \texttt{R:icrf} package.

\section{Simulation studies} 
\label{sec3}

\begin{table}[t!]
    \centering
    \caption{The simulation results for both partly interval-censored (PIC) and fully interval-censored (IC) data under the mild heteroskedasticity  scenario (M1) with $n=200$.}
    \medskip 
    \small
    \begin{tabular}{ccccrrrrcrrrrr}
    \hline
    &&&&\multicolumn{4}{c}{ICQR-KS} && \multicolumn{4}{c}{ICQR-RF}\\
    \cline{5-8}\cline{10-13}
    $\tau$&Error & Cens & Par & Bias & ESE & BSE & CP &&
    Bias & ESE & BSE & CP & RE\\
    \hline\hline
0.3&EV & PIC & $\beta_0$&--0.011 & 0.176 & 0.183 & 0.941 &  & 0.007 & 0.170 & 0.180 & 0.943 & 1.074 \\ 
  &&& $\beta_1$&0.003 & 0.208 & 0.221 & 0.945 &  & --0.009 & 0.206 & 0.224 & 0.962 & 1.018 \\ 
  &&& $\beta_2$&--0.007 & 0.223 & 0.239 & 0.955 &  & --0.016 & 0.223 & 0.240 & 0.962 & 0.996 \\ 
  & & IC & $\beta_0$&--0.067 & 0.184 & 0.192 & 0.920 &  & --0.004 & 0.168 & 0.181 & 0.952 & 1.358 \\ 
  &&& $\beta_1$&0.032 & 0.213 & 0.228 & 0.951 &  & --0.013 & 0.209 & 0.225 & 0.951 & 1.058 \\ 
  &&& $\beta_2$&0.024 & 0.228 & 0.245 & 0.953 &  & --0.014 & 0.223 & 0.242 & 0.958 & 1.053 \\ 
  &Logis & PIC & $\beta_0$&--0.005 & 0.306 & 0.318 & 0.940 &  & 0.009 & 0.297 & 0.314 & 0.950 & 1.061 \\ 
  &&& $\beta_1$&--0.071 & 0.348 & 0.374 & 0.939 &  & --0.069 & 0.354 & 0.382 & 0.945 & 0.970 \\ 
  &&& $\beta_2$&--0.013 & 0.385 & 0.415 & 0.958 &  & --0.018 & 0.385 & 0.417 & 0.959 & 0.999 \\ 
  & & IC & $\beta_0$&--0.055 & 0.316 & 0.325 & 0.922 &  & --0.001 & 0.296 & 0.313 & 0.946 & 1.174 \\ 
  &&& $\beta_1$&--0.045 & 0.357 & 0.377 & 0.942 &  & --0.074 & 0.346 & 0.380 & 0.946 & 1.034 \\ 
  &&& $\beta_2$&0.008 & 0.394 & 0.425 & 0.965 &  & --0.017 & 0.382 & 0.417 & 0.962 & 1.062 \\ 
  &Chi & PIC & $\beta_0$&0.008 & 0.263 & 0.280 & 0.958 &  & 0.021 & 0.257 & 0.277 & 0.960 & 1.041 \\ 
  &&& $\beta_1$&--0.047 & 0.313 & 0.330 & 0.950 &  & --0.053 & 0.315 & 0.336 & 0.955 & 0.982 \\ 
  &&& $\beta_2$&--0.033 & 0.334 & 0.356 & 0.956 &  & --0.039 & 0.332 & 0.358 & 0.961 & 1.008 \\ 
  & & IC & $\beta_0$&--0.048 & 0.277 & 0.290 & 0.940 &  & 0.010 & 0.258 & 0.278 & 0.956 & 1.186 \\ 
  &&& $\beta_1$&--0.023 & 0.318 & 0.334 & 0.949 &  & --0.057 & 0.311 & 0.335 & 0.961 & 1.017 \\ 
  &&& $\beta_2$&--0.006 & 0.344 & 0.363 & 0.945 &  & --0.037 & 0.336 & 0.359 & 0.959 & 1.036 \\ [3pt]
  
  0.5&EV & PIC & $\beta_0$&--0.001 & 0.200 & 0.208 & 0.942 &  & 0.011 & 0.196 & 0.207 & 0.948 & 1.038 \\ 
  &&& $\beta_1$&--0.004 & 0.244 & 0.251 & 0.941 &  & --0.008 & 0.245 & 0.254 & 0.950 & 0.991 \\ 
  &&& $\beta_2$&--0.013 & 0.267 & 0.272 & 0.951 &  & --0.022 & 0.265 & 0.274 & 0.951 & 1.011 \\ 
  & & IC & $\beta_0$&--0.056 & 0.210 & 0.216 & 0.921 &  & 0.014 & 0.194 & 0.203 & 0.955 & 1.249 \\ 
  &&& $\beta_1$&0.026 & 0.252 & 0.255 & 0.936 &  & --0.017 & 0.243 & 0.250 & 0.943 & 1.082 \\ 
  &&& $\beta_2$&0.016 & 0.275 & 0.278 & 0.944 &  & --0.029 & 0.266 & 0.272 & 0.948 & 1.060 \\ 
  &Logis & PIC & $\beta_0$&--0.026 & 0.272 & 0.280 & 0.940 &  & --0.012 & 0.266 & 0.279 & 0.950 & 1.053 \\ 
  &&& $\beta_1$&--0.038 & 0.323 & 0.337 & 0.951 &  & --0.042 & 0.325 & 0.345 & 0.952 & 0.985 \\ 
  &&& $\beta_2$&0.008 & 0.360 & 0.369 & 0.941 &  & 0.003 & 0.362 & 0.372 & 0.947 & 0.989 \\ 
  & & IC & $\beta_0$&--0.073 & 0.282 & 0.288 & 0.933 &  & --0.012 & 0.263 & 0.274 & 0.946 & 1.224 \\ 
  &&& $\beta_1$&--0.019 & 0.327 & 0.340 & 0.952 &  & --0.051 & 0.320 & 0.340 & 0.957 & 1.022 \\ 
  &&& $\beta_2$&0.033 & 0.367 & 0.376 & 0.941 &  & --0.001 & 0.361 & 0.369 & 0.947 & 1.042 \\ 
  &Chi & PIC & $\beta_0$&0.021 & 0.355 & 0.371 & 0.942 &  & 0.024 & 0.356 & 0.373 & 0.945 & 0.993 \\ 
  &&& $\beta_1$&--0.062 & 0.436 & 0.431 & 0.938 &  & --0.060 & 0.439 & 0.446 & 0.944 & 0.988 \\ 
  &&& $\beta_2$&--0.057 & 0.456 & 0.474 & 0.947 &  & --0.056 & 0.459 & 0.480 & 0.949 & 0.988 \\ 
  & & IC & $\beta_0$&--0.029 & 0.367 & 0.379 & 0.940 &  & 0.022 & 0.350 & 0.367 & 0.940 & 1.102 \\ 
  &&& $\beta_1$&--0.041 & 0.441 & 0.429 & 0.929 &  & --0.067 & 0.433 & 0.434 & 0.941 & 1.022 \\ 
  &&& $\beta_2$&--0.040 & 0.467 & 0.480 & 0.941 &  & --0.061 & 0.454 & 0.476 & 0.948 & 1.047 \\ 

\hline
    \end{tabular}
    \label{tab1}
    
	\begin{tablenotes}
		\item Note: Bias, empirical bias; ESE, empirical standard error; BSE, average of bootstrap standard error;
		CP, coverage probability of 95\% Wald-type bootstrap confidence intervals; RE, relative efficiency of ICQR-RF over ICQR-KS.
	\end{tablenotes}
\end{table}

\begin{table}[t!]
    \centering
    \caption{The simulation results for both partly interval-censored (PIC) and fully interval-censored (IC) data under the moderate heteroskedasticity scenario (M2) with $n=200$.}
    \medskip 
    \small
    \begin{tabular}{ccccrrrrcrrrrr}
    \hline
    &&&&\multicolumn{4}{c}{ICQR-KS} && \multicolumn{4}{c}{ICQR-RF}\\
    \cline{5-8}\cline{10-13}
    $\tau$&Error & Cens & Par & Bias & ESE & BSE & CP &&
    Bias & ESE & BSE & CP & RE\\
    \hline\hline
    0.3&EV & PIC & $\beta_0$&--0.003 & 0.203 & 0.213 & 0.945 &  & 0.010 & 0.199 & 0.211 & 0.941 & 1.038 \\ 
  & & & $\beta_1$&--0.003 & 0.250 & 0.265 & 0.948 &  & --0.011 & 0.251 & 0.268 & 0.954 & 0.990 \\ 
  & & & $\beta_2$&--0.013 & 0.250 & 0.267 & 0.960 &  & --0.019 & 0.250 & 0.270 & 0.962 & 0.997 \\ 
& & IC& $\beta_0$&--0.059 & 0.213 & 0.221 & 0.920 &  & --0.003 & 0.200 & 0.211 & 0.945 & 1.221 \\ 
  & & & $\beta_1$&0.029 & 0.259 & 0.270 & 0.943 &  & --0.012 & 0.251 & 0.268 & 0.952 & 1.076 \\ 
  & & & $\beta_2$&0.014 & 0.254 & 0.273 & 0.962 &  & --0.017 & 0.251 & 0.272 & 0.961 & 1.022 \\ 
  
   &Logis & PIC& $\beta_0$& 0.009 & 0.354 & 0.366 & 0.940 &  & 0.016 & 0.349 & 0.365 & 0.945 & 1.027   \\
  &&& $\beta_1$& --0.088 & 0.420 & 0.442 & 0.940 &  & --0.086 & 0.422 & 0.452 & 0.945 & 0.993  \\ 
  &&& $\beta_2$&  --0.021 & 0.431 & 0.461 & 0.959 &  & --0.021 & 0.427 & 0.465 & 0.958 & 1.019  \\   
  &&IC& $\beta_0$& --0.039 & 0.364 & 0.370 & 0.927 &  & 0.004 & 0.346 & 0.360 & 0.947 & 1.119    \\   
  &&& $\beta_1$& --0.073 & 0.424 & 0.437 & 0.936 &  & --0.092 & 0.419 & 0.445 & 0.936 & 1.006 \\ 
  &&& $\beta_2$& 0.001 & 0.440 & 0.472 & 0.965 &  & --0.019 & 0.431 & 0.465 & 0.957 & 1.040   \\    
  
  &Chi & PIC& $\beta_0$&  0.018 & 0.306 & 0.323 & 0.957 &  & 0.025 & 0.304 & 0.322 & 0.960 & 1.010   \\ 
  &&& $\beta_1$& -0.056 & 0.377 & 0.391 & 0.947 &  & --0.058 & 0.377 & 0.399 & 0.950 & 0.998     \\ 
  &&& $\beta_2$& -0.039 & 0.376 & 0.394 & 0.950 &  & --0.042 & 0.373 & 0.398 & 0.954 & 1.014  \\   
  &&IC & $\beta_0$& --0.033 & 0.318 & 0.331 & 0.945 &  & 0.017 & 0.301 & 0.321 & 0.959 & 1.125     \\ 
  &&& $\beta_1$& --0.031 & 0.378 & 0.393 & 0.942 &  & --0.063 & 0.374 & 0.396 & 0.952 & 1.000   \\ 
  &&& $\beta_2$& --0.023 & 0.380 & 0.401 & 0.951 &  & --0.044 & 0.374 & 0.400 & 0.957 & 1.022 \\ [3pt]

0.5&EV & PIC & $\beta_0$&0.005 & 0.228 & 0.243 & 0.952 &  & 0.014 & 0.227 & 0.241 & 0.955 & 1.005 \\ 
  & & & $\beta_1$&--0.006 & 0.291 & 0.300 & 0.946 &  & --0.012 & 0.291 & 0.303 & 0.948 & 0.999 \\ 
  & & & $\beta_2$&--0.018 & 0.296 & 0.305 & 0.950 &  & --0.023 & 0.297 & 0.307 & 0.949 & 0.991 \\ 
  & & IC& $\beta_0$&--0.047 & 0.238 & 0.250 & 0.932 &  & 0.013 & 0.223 & 0.237 & 0.948 & 1.179 \\ 
  & & & $\beta_1$&0.020 & 0.298 & 0.304 & 0.942 &  & --0.018 & 0.289 & 0.298 & 0.951 & 1.064 \\ 
  & & & $\beta_2$&0.006 & 0.306 & 0.310 & 0.944 &  & --0.025 & 0.297 & 0.305 & 0.946 & 1.054 \\ 
  
  & Logis & PIC & $\beta_0$& --0.017 & 0.314 & 0.324 & 0.946 &  & --0.006 & 0.311 & 0.324 & 0.946 & 1.022  \\
  &&& $\beta_1$& --0.045 & 0.386 & 0.400 & 0.952 &  & --0.049 & 0.388 & 0.409 & 0.950 & 0.987  \\ 
  &&& $\beta_2$& 0.002 & 0.399 & 0.411 & 0.938 &  & --0.001 & 0.403 & 0.416 & 0.940 & 0.980    \\   
  && IC & $\beta_0$& --0.064 & 0.325 & 0.326 & 0.935 &  & --0.012 & 0.306 & 0.316 & 0.948 & 1.170       \\   
  &&& $\beta_1$& --0.028 & 0.387 & 0.398 & 0.947 &  & --0.056 & 0.380 & 0.400 & 0.950 & 1.020 \\ 
  &&& $\beta_2$&  0.021 & 0.408 & 0.418 & 0.945 &  & --0.001 & 0.402 & 0.412 & 0.949 & 1.033 \\   
  
  & Chi & PIC & $\beta_0$& 0.034 & 0.411 & 0.427 & 0.936 &  & 0.029 & 0.415 & 0.432 & 0.943 & 0.983   \\ 
  &&& $\beta_1$&  --0.069 & 0.516 & 0.504 & 0.922 &  & --0.064 & 0.521 & 0.522 & 0.934 & 0.984    \\ 
  &&& $\beta_2$&   --0.065 & 0.508 & 0.522 & 0.943 &  & --0.063 & 0.511 & 0.529 & 0.952 & 0.989 \\   
  && IC & $\beta_0$& --0.013 & 0.417 & 0.425 & 0.934 &  & 0.038 & 0.403 & 0.419 & 0.939 & 1.062     \\ 
  &&& $\beta_1$& --0.062 & 0.510 & 0.489 & 0.913 &  & --0.076 & 0.506 & 0.503 & 0.934 & 1.008      \\ 
  &&& $\beta_2$&  --0.054 & 0.516 & 0.524 & 0.937 &  & --0.075 & 0.503 & 0.525 & 0.950 & 1.041   \\

\hline
    \end{tabular}
    \label{tab2}
    
	\begin{tablenotes}
		\item Note: See the note of Table \ref{tab1}.
	\end{tablenotes}
\end{table}

In this section, we conduct extensive Monte-Carlo simulation studies to verify the finite-sample performance of the proposed method for various types of interval-censored data.
All simulations involve two covariates,
$\xx_i=(x_{1i},x_{2i})$, where 
$x_{1i} \sim \text{Uniform}(-1,1)$ and $x_{2i} \sim \text{Bernoulli}(0.5)$. Then, the log-transformed failure time $T_i$ for the $i$th subject  is generated from
$$
T_i = \beta_0 + \beta_1(\tau)x_{1i} + \beta_2(\tau)x_{2i} + \sigma(\xx_i) e_i(\tau),~i=1,\ldots,n, 
$$
where $\beta_0=1.5$, $\beta_1(\tau)=\beta_2(\tau) = 1 $, and
$e_i(\tau) = \varepsilon_i - F_\varepsilon^{-1} (\tau)$ is a quantile-specific random error variable with cumulative distribution function $F_\varepsilon$. 
The random error $\varepsilon_i$ follows one of three distributions: 
(i) extreme value  distribution with location and scale parameters $-1$ and 1, respectively (``EV''),
(ii) logistic distribution with location and scale parameters $-2$ and 1, respectively (``Logis''), and
(iii) chi-squared distribution with degree of freedom 3 (``Chi'').
The degree of error heteroskedasticity is determined by letting (M1): $\sigma(\xx)=1 + 0.3(1-x_1)^2$ and (M2): $\sigma(\xx)=1 + 0.5(1-x_1)^2$.
Note that (M2) produces more pronounced heteroskedasticity than (M1). 
The sample size is $n=200$ or $400$ and 
the quantile level is set to $\tau=0.3$ and 0.5.
All simulation results are summarized from 1000 data replicates with 200 bootstrap samples.

We examine the proposed interval-censored quantile regression (ICQR) estimators under partly interval-censoring (PIC) and fully interval-censoring (IC) sampling schemes.
For the $i$th subject, the censoring time $C_i$ is generated as  $e^{C_i}\sim\text{Uniform}(30,50)$. Then a sequence of $K_i$ examination times $(U_{i1},\ldots,U_{iK_i})$ is created from
$e^{U_{ik}} = e^{U_{i(k-1)}}+\text{Uniform}(0.1,1)  , ~k=1,\ldots,K_i,$ where $K_i$ is the largest integer that satisfies  ${U_{iK_i}} \le C_i$, and the interval $(L_i,R_i)$ that contains $T_i$ is created by 
$L_i=\max_k\{U_{ik}: U_{ik} \le T_i\}$ and 
$R_i=\min_k\{U_{ik}: U_{ik} > T_i\}) $.
To mix exact and interval-censored data, we generate $\Delta_i$ from  $P(\Delta_i=1|\xx_i)=p_0-0.1x_{2i}$
and let $T_i$ be observed if $T_i < C_i$, { for some $p_0\in(0,1)$ that yields approximately 50\% censoring.}
The effective PIC data for subject $i$ are then $\{\Delta_iT_i,(1-\Delta_i)L_i,(1-\Delta_i)R_i,\xx_i\}$.
Letting $\Delta_i=0$ for all subjects reduces 
the PIC data to the IC data.

\begin{figure}[t!]
    \centering
    \includegraphics[width=0.65\textwidth]{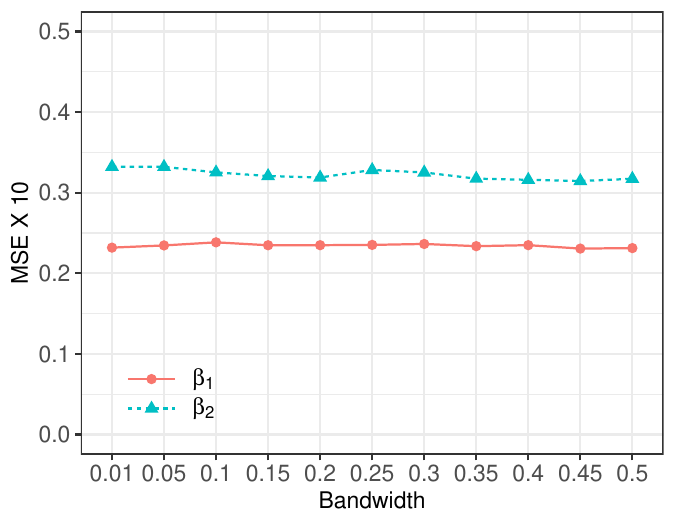}
    \caption{Sensitivity analysis of bandwidth selection when the kernel-based weighting method (``ICQR-KS'') is employed. Red circle and blue triangle points represent mean squared error (MSE) multiplied by 10 for estimating $\beta_1$ and $\beta_2$, respectively, and blue square points represents integrated mean sqaured error for estimating $F(\cdot|\xx)$.}
    \label{fig1}
\end{figure}

Tables \ref{tab1} and \ref{tab2} summarize the performance of the proposed ICQR estimators in the simulations, 
where kernel smoothing (``ICQR-KS'') and random forests (``ICQR-RF'') are used for the weight computation, under (M1) and (M2)  scenarios, respectively. 
Four criteria are used for estimation evaluation:
empirical bias (Bias), empirical standard error (ESE), average of bootstrap standard error (BSE),
and coverage probability (CP) of 95\% bootstrap confidence intervals.
The perturbed resampling and classical percentile bootstrap methods are used for standard error estimation. 
The results demonstrate that the proposed approach is satisfactory with negligible biases, regardless of the magnitude of heteroskedasticity.
The standard error estimators are almost similar to the empirical standard errors, and the coverage probabilities {approximate the nominal value substantially}, implying that the proposed inferential procedure works {effectively} under general interval-censoring sampling schemes.

To compare two weighting methods, the relative efficiency (RE) of ICQR-RF over ICQR-KS in terms of mean sqaured error (MSE) is also calculated. 
In the (M1) scenario, ICQR-RF outperforms ICQR-KS, achieving up to about 25\% efficiency improvement.
In the (M2) scenario, on the other hand, 
ICQR-KS gives slightly more efficient results than  ICQR-RF.
To explore how the proposed kernel-based quantile estimator is robust to bandwidth selection, we also conduct a sensitivity analysis by changing the bandwidth from 0.01 to 0.5 and accordingly computing the MSE of the ICQR-KS estimator.
Figure \ref{fig1} illustrates that the MSE curves remain very stable across a range of bandwidths, suggesting that the selection of an optimal bandwidth might not be a significant issue in the analysis of censored quantile regression.
However, we found that it is important to the estimation of the error distribution $F(\cdot)$. The related simulation results are detailed in the web-appendix.

\begin{figure}[t!]
	\centering
 \includegraphics[width=1\textwidth]{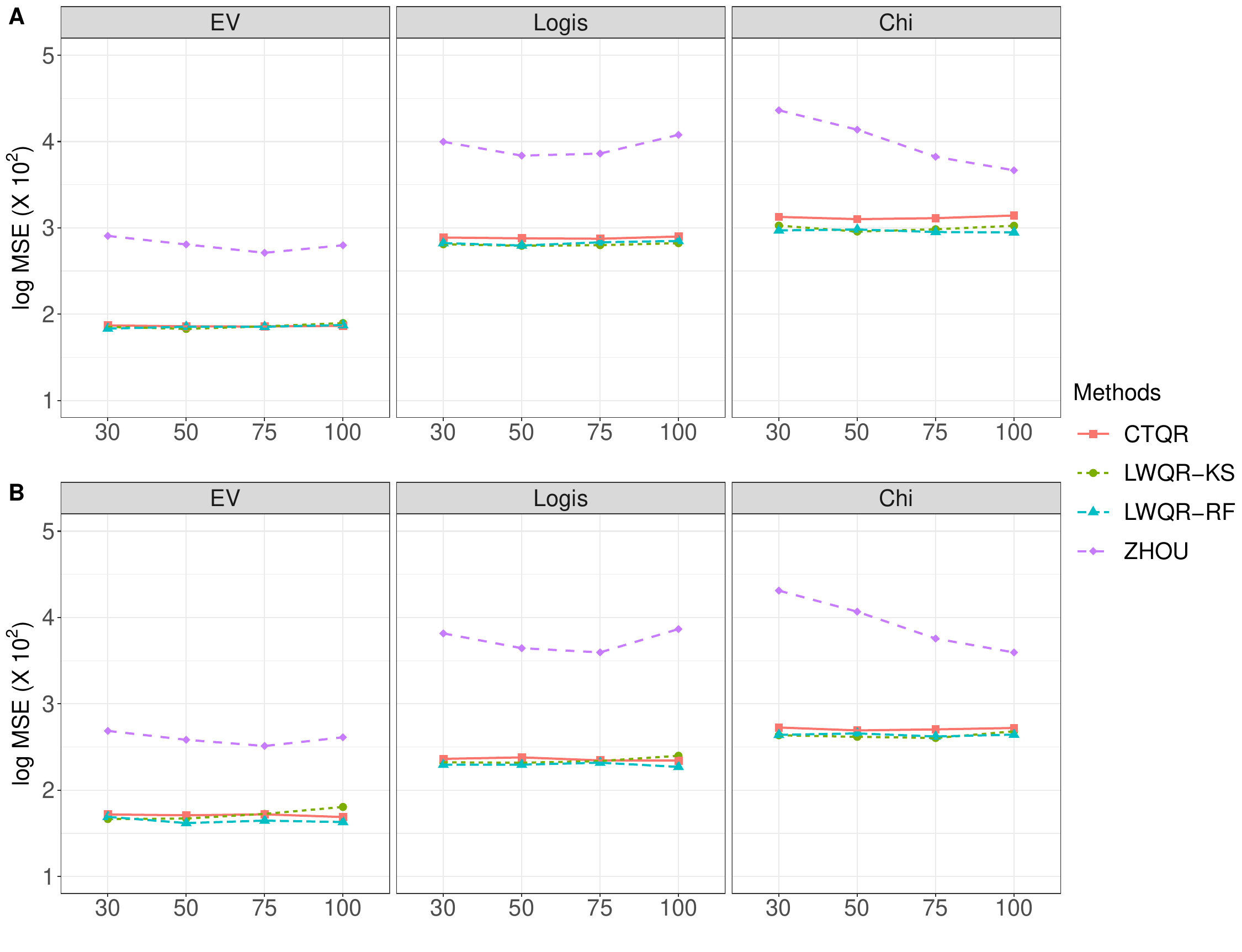}
	\caption{
	Comparison of four interval-censored quantile regression estimators (ICQR-KS, ICQR-RF, CTQR and ZFD). Model performance is measured in terms of the log-transformed MSE across two distinct sample sizes (A: $n=200$ and B: $n=400$) over a spectrum of censoring rates spanning from 30\% to 100\%.}\label{fig2}
\end{figure}

Next, our ICQR methods are compared with two existing inferential procedures: \cite{ctqr,fru22} (``CTQR'') and \cite{zh17} (``ZFD''). 
The ZFD estimator is expected to be statistically inefficient because it does not use the entire data information, while the CTQR method specifies the error distribution parametrically and thus is possibly subject to model mis-specification issues.
Figure~\ref{fig2} displays the log-transformed MSEs 
($\log \{E(\|\hat\bbeta-\bbeta_0\|^2)\times10^2\}$)  
of the four estimation methods. 
Here, three random error distributions (EV, Logis, and Chi) with different sample sizes (A: $n=200$; B: $n=400$) are considered under the (M2) condition. 
In every scenario, ICQR-KS and ICQR-RF consistently outperform other methods, while ZFD exhibits notably inferior performance as anticipated.
Generally, CTQR demonstrates strong performance when dealing with relatively mild asymmetry in the error term (EV and Logis), but exhibits comparatively higher MSE when confronted with asymmetric error terms (Chi).
Notably, the proposed quantile estimators are less sensitive to the censoring rate. 
Overall, our method performs well without a noticeable loss of efficiency even when all event times are interval-censored. This may imply that our regression estimators can retain the $n^{1/2}$-convergence rate, at least empirically, for interval-censored data, even if the nonparametric estimator $\hat F(\cdot|\xx)$ for $F(\cdot|\xx)$ has a much lower convergence rate than $n^{1/2}$.

\section{Two data examples}
\label{sec4}

\subsection{HIV-infected drug user data}

During 1995-1996, 1209 HIV-seronegative injecting drug users were recruited from Bangkok Metropolitan Administration to participate in a prospective cohort study \citep{van01}. 
The study objective was to determine changes in risk behavior in relation to study participation among the drug users. 
Clinic visits occurred every 4 months, at which data were collected on demographics and risk behaviors since the last visit.
However, because the time of seroconversion was known to be only between the last seronegative test and the first seropositive test, the event times were possibly interval-censored. 
As of December 1998, there were 133 HIV-1 seroconversions and approximately 2300 person-years of follow-up. 
Since HIV incidence may vary over time, the appropriate time scale for this analysis was calendar time starting at the earliest enrollment date.

\begin{figure}[t!]
	\centering
	\includegraphics[width=0.7\textwidth]{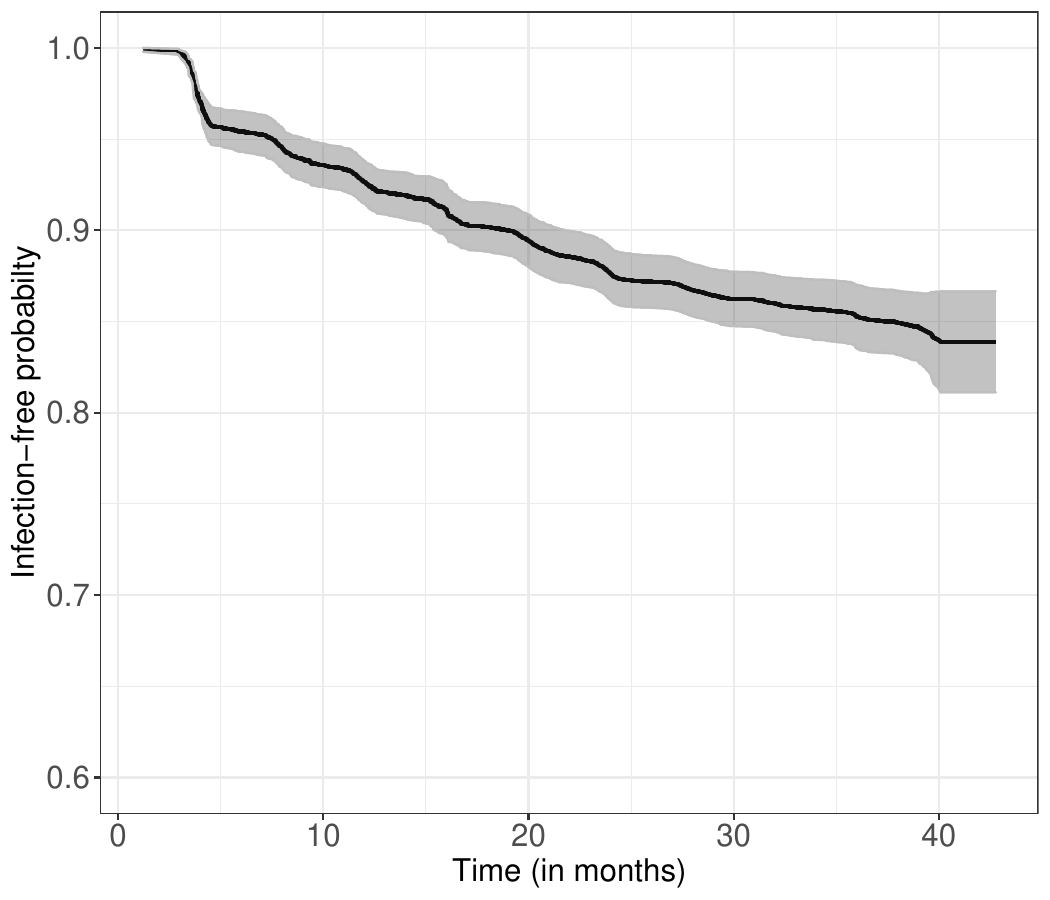}
	\caption{Estimated infection-free probability curve along with 95\% pointwise confidence intervals.}
	\label{fig3}
\end{figure}

Our analysis was based on 1124 individuals,
who visited a clinic at least once, with five covariates:
age at recruitment, gender (0, female; 1, male), 
whether they had needle sharing (0, no; 1, yes),  imprisoned (0, no; 1, yes), and drug injection in jail or not (0, no; 1, yes).
We aim to evaluate the quantile effects of these risk factors on log-transformed time-to-HIV-seropositivity.
Figure \ref{fig3} displays the estimated infection-free probability
curve, along with the 95\% pointwise confidence interval, computed from the self-consistency equation \citep{tu76}. 
Since survival probabilities are all above 80\%, we focus on two quantile levels, $\tau=0.15$ and $0.3$, for our illustration. 
Three interval-censored quantile regression methods, specifically (i) ICQR-KS, (ii) ICQR-RF, and (iii) CTQR, are employed on this dataset, and their results are presented in Table \ref{tab_res}.
Each method is evaluated with regression parameter estimator (Est), boostrap standard error estimate (SE), and lower bound (LB) and upper bound (UB) of the 95\% Wald type confidence interval.

\begin{table}[t!]
	\centering
	\caption{Quantile regression analysis of interval-censored HIV-infected drug user data.}
	\label{tab_res}
 \medskip 
	\begin{tabular}{ccrrrrcrrrr}
		\toprule
		&&\multicolumn{4}{c}{$\tau=0.15$}&&\multicolumn{4}{c}{$\tau=0.3$}\\
		\cmidrule{3-6}\cmidrule{8-11}
		\multicolumn{1}{c}{Variable} & \multicolumn{1}{c}{Method} & 
		\multicolumn{1}{c}{Est} & \multicolumn{1}{c}{SE} & 
		\multicolumn{1}{c}{LB} & \multicolumn{1}{c}{UB} && 
		\multicolumn{1}{c}{Est} & \multicolumn{1}{c}{SE} & 
		\multicolumn{1}{c}{LB} & \multicolumn{1}{c}{UB} 
		\\\midrule
		Intercept&ICQR-KS&1.046 & 0.407 & 0.249 & 1.843 &&2.522 & 0.186 & 2.158 & 2.886 \\ 
		&ICQR-RF&2.123 & 0.426 & 1.288 & 2.958 &&2.936 & 0.378 & 2.194 & 3.678 \\ 
		&CTQR&0.541 & 2.288 & --3.944 & 5.025 &&3.843 & 1.980 & --0.038 & 7.725 \\ [3pt]
		
		Age&ICQR-KS&0.040 & 0.009 & 0.022 & 0.058&&0.015 & 0.005 & 0.005 & 0.024 \\ 
		&ICQR-RF& 0.027 & 0.011 & 0.004 & 0.049 &&0.018 & 0.017 & --0.015 & 0.052 \\ 
		&CTQR&0.084 & 0.064 & --0.040 & 0.209 &&0.008 & 0.053 & --0.095 & 0.111 \\ [3pt]
		
		Gender&ICQR-KS&--0.122 & 0.248 & --0.609 & 0.365&&--0.007 & 0.075 & --0.153 & 0.140 \\
		&ICQR-RF&--0.092 & 0.157 & --0.400 & 0.216 &&0.116 & 0.296 & --0.464 & 0.697 \\ 
		&CTQR&1.323 & 1.963 & --2.526 & 5.171 &&0.617 & 1.125 & --1.588 & 2.823 \\ [3pt]
		
		Needle&ICQR-KS&--0.083 & 0.216 & --0.507 & 0.341&&--0.029 & 0.101 & --0.228 & 0.169 \\ 
		&ICQR-RF&--0.037 & 0.154 & --0.339 & 0.266 &&--0.083 & 0.135 & --0.348 & 0.181 \\ 
		&CTQR&0.056 & 1.071 & --2.044 & 2.155 &&0.126 & 1.813 & --3.428 & 3.680 \\[3pt]
		
		Imprison&ICQR-KS&--0.026 & 0.249 & --0.514 & 0.463&&--0.035 & 0.090 & --0.212 & 0.142 \\ 
		&ICQR-RF&  --0.251 & 0.166 & --0.576 & 0.075&&--0.133 & 0.209 & --0.543 & 0.277 \\ 
		&CTQR&--0.719 & 0.801 & --2.288 & 0.850 &&--0.335 & 0.898 & --2.095 & 1.425 \\ [3pt]
		
		Inject&ICQR-KS&0.234 & 0.172 & --0.103 & 0.572 &&0.084 & 0.068 & --0.050 & 0.217 \\ 
		&ICQR-RF& 0.139 & 0.106 & --0.069 & 0.347&&0.002 & 0.238 & --0.465 & 0.469 \\ 
		&CTQR&0.373 & 1.085 & --1.754 & 2.499 &&0.195 & 1.296 & --2.345 & 2.736 \\
		\bottomrule
	\end{tabular}
\end{table}

Overall, ICQR-KS and ICQR-RF provide very similar estimation results at the two quantile levels, whereas CTQR has slightly different estimates with highly inflated standard errors.
According to the proposed methods, aging is the only significant factor that reduces the risk of HIV-1 infection and prolongs time-to-seropositivity. 
However, its effect size decreases as the quantile level increases from $\tau=0.15$ to $\tau=0.3$, meaning that older people are more likely to be susceptible to early HIV infection.
Gender, needle sharing, imprisonment and drug injection are not significant factors to time-to-seropositivity.
However, it is difficult to generalize these results because we focused on the early stages of events due to the high rate of censorship.

\subsection{HIV-infected children data}

Next, we consider the doubly-censored AIDS clinical trial data collected in a phase-II randomized pediatric clinical trial in 1997 \citep{na00}.
The dataset consists of 298 HIV-infected children to compare the effects of three treatments on the viral response of the HIV virus: (A) zidovudine plus lamivudine, (B) stavudine plus ritonavir, and (C) zidovudine plus lamivudine plus ritonavir.
Previous mean-based analyses \citep{ch20,ch21} showed that treatments A and B do not differ significantly, so they are merged into a single treatment (A/B, coded {as} 0) and compared to treatment C (coded {as} 1).
In general, the effectiveness of HIV treatment is determined by the HIV virus load (i.e., viral RNA copy/ml), where a high viral load may imply deterioration of the disease due to HIV infection or ineffective treatment of HIV.

The study's objective is to evaluate the efficacy of different treatment options in diminishing HIV viral load. The primary measure for comparison is the logarithmic change in HIV viral RNA levels from an initial baseline value ($l_0=\log\text{RNA}_0$) and the measurement after 24 weeks ($l_{24}=\log\text{RNA}_{24}$).
However, it is known that the measurement of RNA assay is reliable only when the values are between $l_0 - 5.88$ and $ l_0-2.6 $.
This limitation results in doubly-censored data, consisting of 52\% exact, 4\% left-censored and 46\% right-censored observations. We then fit the linear quantile regression model to this dataset
\begin{equation*}\label{aideq}
	\log\text{RNA}_{24} =\beta_0(\tau)+\beta_1(\tau)\text{Treatment} +\beta_2(\tau)\log\text{RNA}_{0} + e(\tau).
\end{equation*}

\begin{figure}[t!]
	\centering
	\includegraphics[width=0.8\textwidth]{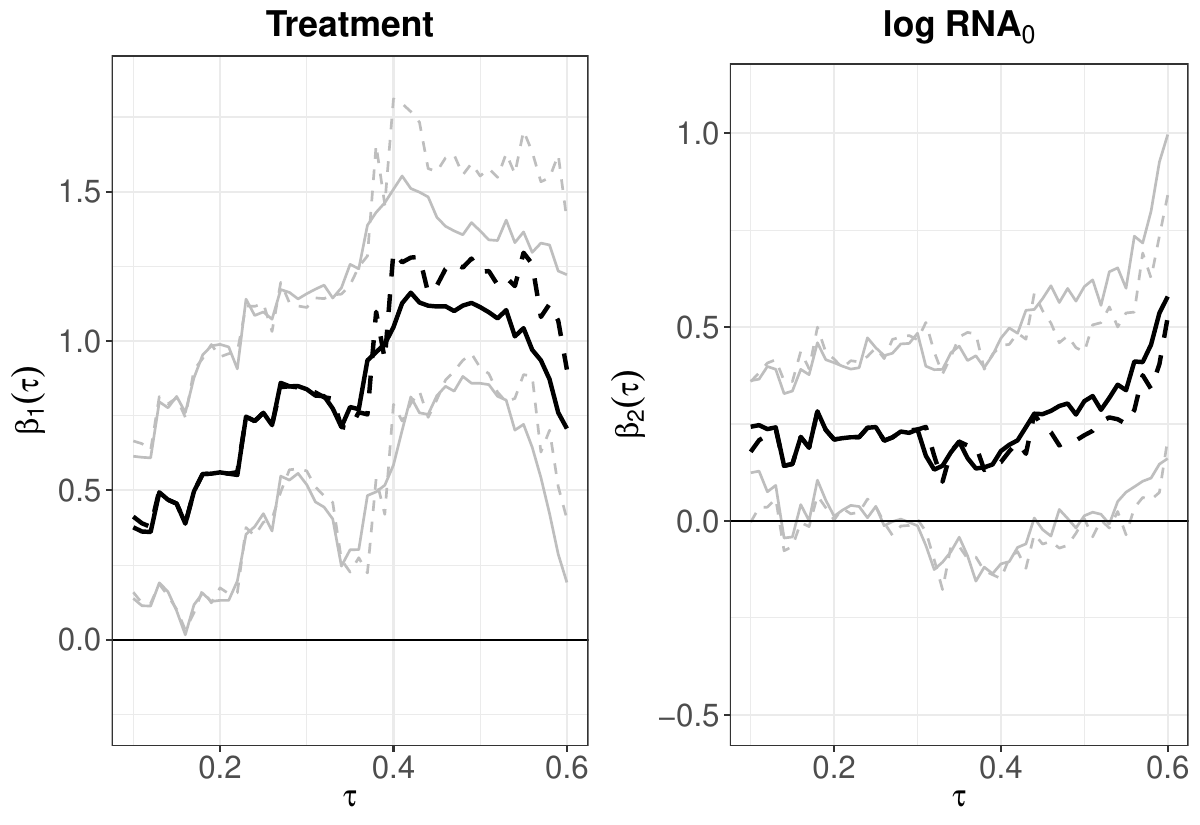}
	\caption{Quantile estimators for doubly-censored HIV-infected children data;
		point estimates with two weighting methods, 
		ICQR-KS (solid line) and ICQR-RF (dashed line),
		along with their 95\% pointwise bootstrapped confidence intervals.}
	\label{fig4}
\end{figure}

Figure \ref{fig4} shows the quantile  effects 
when $\tau$ ranges from 0.1 to 0.5, where the ICQR-KS and ICQR-RF estimators are represented by  solid and dashed lines, respectively, along with their common 95\% pointwise confidence intervals. 
It shows that after accounting for  the baseline RNA level, treatment C significantly increases the risk of HIV virus infection more than treatment A/B after 24 weeks at the quantile level $\tau\in[0.1,0.6]$.
This means that A/B is more effective in controlling the amount of virus in HIV-infected children.

\section{Discussion} 
\label{sec5}

The quantile regression model has been widely studied in statistics and econometrics. The development of appropriate methodologies, fast algorithms, and solid statistical theory under complex censoring schemes is an important but challenging task. 
In this article, we have developed a locally weighted quantile regression method for analyzing general interval-censored data, where the weights are computed with (i) nonparametric kernel smoothing  and (ii) random forests.
Our estimators make effective use of the information carried in the original data and thus are more statistically efficient than existing methods. 
The proposed nonparametric adjustments  may need a tuning parameter, but fortunately it is less sensitive in quantile regression and the rule-of-thumb method would suffice for practical use. 
 We have demonstrated  via extensive numerical studies that our methods are more appealing than existing methods, because they are not only robust to the distribution of the random errors but also in many cases lead to comparable or
smaller MSEs in the estimation of the model parameters.

It should be noted that the proposed estimating equation  can also be derived from the following self-consistency equation \citep{tu76,hu99} of the form
\begin{equation*} \label{se}
\frac1n\sum_{i=1}^n\Delta_i \hat F_n(t) + \frac1n  \sum_{i=1}^n (1-\Delta_i)\frac{\hat F_n(R_i\wedge t)-\hat F_n(L_i\wedge t)}{\hat F_n(R_i)-\hat F_n(L_i)} = \hat F_n(t),
\end{equation*}
where $\hat F_n(\cdot)$ denotes the self-consistent estimator for $F(\cdot)$ with partly interval-censored data.
Since $F(R_i\wedge t)-F(L_i\wedge t)=\{F(R_i)-F(t)\}I(R_i<t) + \{F(t)-F(L_i)\}I(L_i<t)$ for a non-decreasing function $F$, the above formula implies that 
\begin{equation*}
\begin{split}
\frac1n\sum_{i=1}^n &  \Delta_i I(T_i<t) + \frac1n  \sum_{i=1}^n (1-\Delta_i)\times \\
&~~\left[ \left\{\frac{\tau -\hat F(L_i)}{\hat F(R_i)-\hat F(L_i)}\right\}I(L_i<t)
+ \left\{ \frac{\hat F(R_i)-\tau}{ \hat F(R_i)- \hat F(L_i)}\right\}I(R_i<t)  \right]= \tau 
\end{split}
\end{equation*}
at the quantile level $\tau$, which  is equivalent to the estimating function in \eqref{ee}. This argument can also be used to show the consistency of the proposed method. 
See also \cite{hu99} for asymptotic properties of nonparametric estimation under general interval-censoring schemes.

In this article, we have established the asymptotic consistency of the proposed quantile regression estimator and leave the weak convergence result as future work. 
The latter requires showing that  $\hat F(\cdot|\xx)$ converges to $F(\cdot|\xx)$ not too slowly, which is, however, rarely known in the literature under the interval-censoring setting. 
Our simulation study showed that the proposed method {performs markedly well}, with minimal loss of efficiency, even in the absence of exact failure time information. This may imply that our regression estimator $\hat\bbeta$ can still achieve the $n^{1/2}$-convergence rate under regular conditions.

We further point out some possible research issues below.
To overcome the curse-of-dimensionality of nonparametric  kernel smoothing, an efficient dimension reduction procedure for interval-censored data is necessary, which can also be used for variable selection  as in regularized censored quantile regression methods \citep{wa13,son22}.  
Recently, \cite{de2019adapted} developed a censored quantile regression method with adaptive quantile loss function.
Their adaptive-loss approach is expected to retain the same level of efficiency with locally weighted procedures as it utilizes full data. 
This method might be computationally advantageous since it is not necessary to solve a complex self-consistency equation for interval-censoring. 
We leave these important and interesting topics for future research.

\section*{Acknowledgements}

The research of T. Choi was supported by the National Research Foundation of Korea (NRF) grant funded by the Ministry of Education (RS-2023-00237435).
The research of S. Choi was supported by the National Research Foundation of Korea (NRF) grant funded by the Korean government (2022M3J6A1063595, 2022R1A2C1008514).

\bibliographystyle{apalike}
\bibliography{biblist}

\end{document}